\documentclass[reprint,
superscriptaddress,
 amsmath,amssymb,
floatfix,
]{revtex4-2}

\usepackage{graphicx}
\usepackage{dcolumn}
\usepackage{bm}
\usepackage[breaklinks=true,colorlinks=true,linkcolor=blue,urlcolor=blue,citecolor=blue]{hyperref}
\usepackage{epstopdf}
\usepackage{siunitx}

\usepackage{enumitem}
\usepackage{color}
\begin{document}


\title{Identification and minimization of losses in microscaled spin-wave transducers}
\author{Felix Kohl}
\email{fkohl@rptu.de}
\affiliation{Fachbereich Physik and Landesforschungszentrum OPTIMAS,
Rheinland-Pfälzische Technische Universität Kaiserslautern-Landau,
D-67663 Kaiserslautern, Germany}

\author{Björn Heinz}
\affiliation{Fachbereich Physik and Landesforschungszentrum OPTIMAS,
Rheinland-Pfälzische Technische Universität Kaiserslautern-Landau,
D-67663 Kaiserslautern, Germany}

\author{\'Ad\'am Papp}
\affiliation{Faculty of Information Technology and Bionics, Pazmany Peter Catholic University, Budapest, Hungary}
\affiliation{Jedlik Innovation Kft., Budapest, Hungary}

\author{R\'obert Erd\'elyi}
\affiliation{Faculty of Information Technology and Bionics, Pazmany Peter Catholic University, Budapest, Hungary}
\affiliation{Jedlik Innovation Kft., Budapest, Hungary}

\author{Gyorgi Csaba}
\affiliation{Faculty of Information Technology and Bionics, Pazmany Peter Catholic University, Budapest, Hungary}
\affiliation{Jedlik Innovation Kft., Budapest, Hungary}

\author{Philipp Pirro}%
\affiliation{Fachbereich Physik and Landesforschungszentrum OPTIMAS,
Rheinland-Pfälzische Technische Universität Kaiserslautern-Landau,
D-67663 Kaiserslautern, Germany}


\date{May 16, 2025}

\begin{abstract}
Magnonics is a promising platform for integrated radio frequency (rf) devices, leveraging its inherent non-reciprocity and reconfigurability. However, the efficiency of spin-wave transducers driven by rf-currents remains a major challenge. In this study, we systematically investigate a spin-wave transducer composed of micron-sized rf antennas on yttrium iron garnet (YIG) films of different thickness - an ideal testbed for integrated magnonic devices. Using propagating spin-wave spectroscopy and numerical simulations, we analyze spin-wave transmission, identifying key loss mechanisms and improving device efficiency by reducing ohmic resistance. The resulting improvements enable the reduction of insertion loss to below $\SI{10}{dB}$ in microscaled spin-wave transducers. At the same time large non-reciprocity  can be exploited to achieve significant isolation on the microscale.
\end{abstract}

\maketitle

\section{Introduction}

Recent advances in information and communication technology continue to drive the demand for efficient, scalable and versatile radio frequency (rf) devices \cite{BalyshevaRFDemand}. This trend is further accelerated by the emergence of Internet-of-Things (IoT) applications, which require improved data transmission rate and processing, to support the increasing number of interconnected devices \cite{ShafiqueIoT}. Hence, there is a growing need for greater bandwidth in wireless communication systems \cite{Mahon5G}.
At the same time, this demand is driven by the ongoing rollout of 5G standards and the development of 6G technologies, both of which promise higher bandwidth and operation at higher frequencies \cite{Mahon5G}. However, established technologies for integrated analog devices, e.g. based on surface acoustic wave (SAW) and bulk acoustic wave (BAW), face significant challenges at high operation frequencies \cite{DaiOctave, AignerBAW, Delsing2019, Hara2010}. The scalability and efficiency of these devices is limited by increased losses and the increasing complexity of their manufacturing. These limitations drive the search for alternative technological approaches.

Magnonics with its focus on the generation, manipulation and utilization of spin waves is a promising approach to tackle these problems. Altogether, it offers several advantages over established alternatives. First and foremost, the operating frequency range of magnetic devices can be tuned via externally accessible parameters such as the applied magnetic field \cite{Stancil, HanMagnonicDevices}. Second, spin waves provide a readily accessible signal non-reciprocity, which allows such transducers to be used as microwave filters and isolators at the same time. For example in this work, the isolation could be shown to reach up to $\SI{30}{dB}$ across the $\SI{3}{dB}$ bandwidth of the devices under test. In addition, spin waves are inherently low energy excitations \cite{Roadmap2024}, rendering them ideal for IoT applications. They also exhibit intrinsic nonlinearity \cite{Breitbach2024} due to the structure of the underlying equations of motion, which can be exploited for various use cases, including neuromorphic computing \cite{Papp2021, Qi2024}. Overall, magnonics offers a versatile and tunable technological option that can be integrated into existing manufacturing processes.\\
Over the past few years, numerous scientific demonstrations of prototype magnonic devices have been presented, proving their potential to achieve various functionalities \cite{QiRepeater, QiCoupler, Roadmap2024,HanMagnonicDevices}. Despite the potential demonstrated, many of these devices are still either  energy inefficient and bulky \cite{Levchenko, Webb}  or small but lossy, with losses exceeding $\SI{20}{dB}$ \cite{Davidkova}. Most of them also rely on electromagnets to generate the external magnetic fields required for operation \cite{DaiOctave, DuFilter}.  Thus, there is a particular need to unravel the mechanisms limiting the energy efficiency, which is key to lift integrated magnonic devices to a competitive level.\\
In this work, a fundamental building block of magnonic devices, known as spin-wave transducer or spin-wave transmission line is investigated using inductive spectroscopy of propagating spin waves \cite{Bailleul2003, Vanataka,Vlaminck2010, DevolderTimeGate}. Commonly, these structures consist of two symmetrical antennas patterned on a magnetic substrate. The objective of this study is to identify the main causes of inefficiency in such magnonic RF devices to enable the future design of efficient magnonic devices. An improved transducer structure is reported to show losses lower than $\SI{10}{dB}$ representing a significant improvement to the performance.


\section{\label{sec:ExpSetup}Experimental setup}

\subsection{\label{sec:MeasMeth}Measurement method}
For the investigation of different spin-wave transducers, the technique of propagating spin wave spectroscopy (PSWS) is used \cite{Vanataka, Vlaminck2010}. In our experiment the device under investigation is connected to a vector network analyzer (VNA) via commercial microwave probes (Picoprobe, GGB Industries). A schematic representation of the setup with device under test is shown in Fig \ref{fig:ExpSetup} a). To mitigate the influence of cables and probes, a short-open load-through (SOLT) calibration is performed on a commercial calibration substrate prior to the measurements. Signal transmission is characterized using the scattering parameter description of a 2-port network \cite{Pozar} across frequency spectra containing the spin-wave resonances. The spectrum of transmitted spin waves can be assessed in both magnitude and phase. From this phase information, the spin-wave propagation can be characterized, e.g. the group velocity \cite{JiapengXu} and the dispersion relation \cite{Vanataka,DevolderTimeGate}. The external field is aligned parallel to the antennas within the film plane, facilitating measurements in the so-called Damon-Eshbach (DE) configuration, where spin waves propagate perpendicular to the direction of static magnetization. If not stated otherwise, the measurements are performed using an VNA output power of $P=\SI{-25}{dBm}$.

In the following, the primary focus is put on the relative strength of spin-wave transmission, i.e. the efficiency and performance of the transducers, which is assessed using the insertion loss \cite{Pozar} 

\begin{equation}
    IL=-20\log\vert S_{ij}\vert= -\text{mag}(S_{ij}).
\end{equation}
Here $i,j\in\{1,2\}$ and $i\neq j$ are the ports of the network and $S_{ij}$ defined as the ratio of outcoming voltage amplitude $V_j$ at port $j$, and incident voltage amplitude $V_i$ at port $i$, when no voltage amplitude is incident on port $j$.
This metric directly correlates with the magnitude of signal transmission, where lower insertion loss describes more efficient signal transmission.

\subsection{\label{sec:SampleFab}Sample fabrication and design}

\begin{figure*}[htb]
\includegraphics{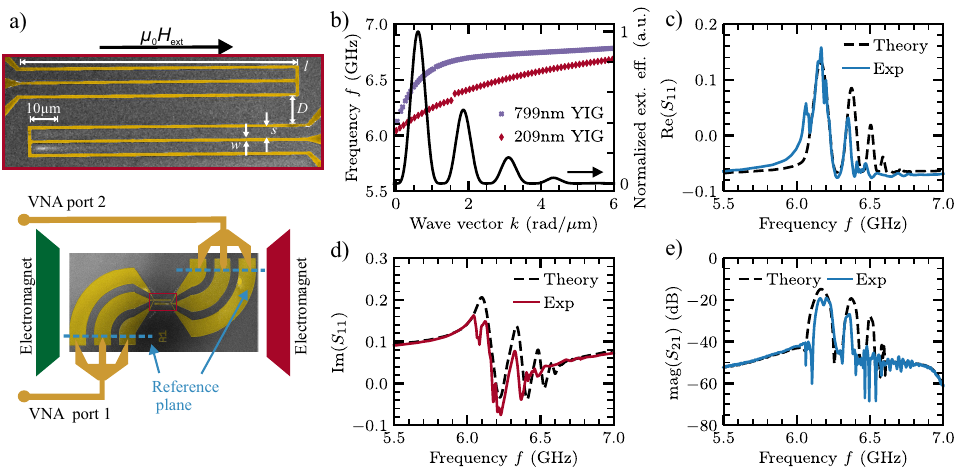}
\caption{a) Schematic representation of the propagating spin-wave spectroscopy setup. The reference plane after SOLT-calibration is indicated. Coplanar waveguide antenna dimensions are indicated in the scanning electron microscope image. They are $l=\SI{100}{\micro\metre}$ long, $D=\SI{10}{\micro\metre}$. The metal lines are $w=\SI{1}{\micro\metre}$ wide and have a spacing of $s=\SI{4}{\micro\metre}$. b) Dispersion relation of the DE spin-wave mode for different YIG thicknesses, calculated by TetraX \cite{TetraX}. The wave-vector spectrum of the coplanar waveguide antenna indicates the parts of the dispersion relation with highest excitation efficiency. Exemplary measured and calculated real (c) and imaginary part (d) of the reflection parameter $S_{11}$ and the magnitude of the transmission parameter $S_{21}$ (e) for $\mu_0 H_\text{ext}=\SI{138.7}{\milli \tesla}$ on the $\SI{209}{\nano\metre}$ YIG film. The electric antenna properties are included according to Section \ref{sec:Theory}, while the calculations are performed according to Eqs. (\ref{equ:SelfInductance}) and (\ref{equ:MutualInductance}).
\label{fig:ExpSetup}}
\end{figure*}

For the spin-wave transducer, two opposing coplanar waveguides (CPWs) with a edge-to-edge spacing of $D=\SI{10}{\micro\metre}$ are patterned onto yttrium iron garnet (YIG) films of varying thicknesses ($\SI{101}{\nano\metre}$, $\SI{209}{\nano\metre}$, $\SI{415}{\nano\metre}$, $\SI{799}{\nano\metre}$), grown on a $\SI{500}{\micro\metre}$ thick (111) gadolinium gallium garnet (GGG) substrate, purchased commercially from the company Matesy GmbH. in Jena, Germany. To facilitate microwave probe contact, the CPWs are connected to curved contact pads. A representative structure is shown in the scanning electron image in Fig. \ref{fig:ExpSetup} a). In addition, to evaluate the exponential spin-wave decay, the antenna spacing was systematically varied from $\SI{10}{\micro\metre}$ to $\SI{500}{\micro\metre}$ (see supplemental material). Fabrication involved electron beam lithography using a PMMA based resist mask, followed by deposition of $\SI{10}{\nano\metre}$ Ti and $\SI{150}{\nano\metre}$ Au using electron beam evaporation and a subsequent lift-off procedure.\\
The wave-vector distribution of the antenna's magnetic field can be interpreted as its excitation efficiency in wave-vector space. In this study, the same antenna design is used for all YIG films.  The design is specifically chosen to maximize excitation efficiency in spin-wave dispersion parts of high group velocity. Consequently, the antenna can only excite fast-propagating spin-wave with large decay length, minimizing losses associated with spin-wave damping, thereby improving overall transmission efficiency. Calculated dispersion relations of the DE mode, using the solver TetraX \cite{Koerber,TetraX}, along with the excitation efficiency of the antenna are shown in \mbox{Fig.\,\ref{fig:ExpSetup}b)}. The gap in dispersion relation for the DE mode of the $\SI{209}{nm}$ thick YIG is caused by an avoided crossing with a higher quantized mode (see Supplemental material).  A more precise modeling of the antennas and spin-wave resistance is carried out using the numerical simulation model from \cite{RobertDesign}. The obtained data can be found in the supplemental material.

\section{\label{sec:Theory} Theoretical background}

\subsection{Analytical description}
While  PSWS is an established technique, the extensive modeling efforts in this field often focus on spin wave characteristics rather than device-level behavior. Frequently, the measured spin wave amplitudes are rather weak, enforcing the use of various post-processing methods such as background subtraction of the electromagnetic background signal \cite{Vanataka, Vlaminck2010} or a timegating approach \cite{DevolderTimeGate}, which are not feasible for self-sufficient rf-devices. If the spin-wave signal is significantly larger than the electromagnetic background, these post-processing procedures are not necessary and the spin-wave signal can be read directly from the acquired signal.
The structures in this work are much smaller than the electromagnetic wavelength \cite{RobertDesign}, which allows for a circuit-level analysis using lumped elements. The response of the magnetic system is accounted for by a radiation impedance term \cite{RobertDesign, Vanderveken} using the simplified analytic description  given in \cite{Vanderveken}. In the following, the essential elements of this analytic model relevant for understanding the results of this work are reproduced here:

We consider an rf-antenna as outlined in Fig. \ref{fig:ExpSetup} a), which provides an oscillating magnetic field around the antenna driving the magnetization in the neighbouring magnetic material. This driven magnetization consequently induces a voltage opposite to the voltage driving the current in the antenna. The induced voltage oscillating at angular frequency $\omega$ can be expressed as function of the total magnetic flux $\Phi$:

\begin{equation}
    V_{\text{ind}} = \oint_{C_\text{a}}\textbf{E}(\textbf{r})d\textbf{l}=-i\omega \int_{S_\text{a}}\textbf{B}(\textbf{r})d^2\textbf{a}= -i\omega \Phi.
\end{equation}

with the surface $S_\text{a}$ enclosing the current loop, $C_\text{a}=\partial S_\text{a}$ and $\textbf{B}$ the magnetic induction.
In this approximation, the total flux $\Phi$ can be separated in the flux from the current itself $\Phi_0=I\cdot L_0$ and the flux caused by the magnetization $\Phi_\text{m}=I\cdot L_\text{m}$.  Here, $L_0$ denotes the antennas self-inductance, while $L_\text{m}$ is the additional inductance introduced by the generation of spin waves. This inductance can be approximately modelled as \ \cite{Vanderveken} 
\begin{equation}
    L_\text{m} = \frac{\Phi_\text{m}}{I}=\mu_0 \int\int \textbf{h}_\text{a}^T(\textbf{r}')\chi_\text{m}(\textbf{r}',\textbf{r})\textbf{h}_\text{a}(\textbf{r})d^3\textbf{r}d^3\textbf{r}'
\end{equation}

using the dynamic magnetic field $\textbf{h}_\text{a}$ and the dynamic susceptibility tensor $\chi_\text{m}$, comprising the spin-wave response.
Assuming uniform magnetization profiles along the magnetic film thickness $t$ and the antenna length $l$, allows to reduce the volume integral to a one dimensional integral, that can be converted to reciprocal space more suited for analytical expression of the dynamic magnetic susceptibility:

\begin{equation}\label{equ:SelfInductance}
    L_\text{m} =\mu_0 tl\int\textbf{h}_a^T(k)\chi \big(\omega(k)\big) \textbf{h}_a^*(k)\frac{dk}{2\pi}
\end{equation}

The same procedure can also be extended to calculate the mutual inductance $\mathcal{M}$ between two antennas mediated by the spin-wave response. The derivation is identical, apart from a phase accumulation due to the spatial separation $D$ introduced by a complex phase factor $kD$ in the integral:

\begin{equation}\label{equ:MutualInductance}
    \mathcal{M} = \mu_0 tl\int\text{e}^{ikD}\textbf{h}_a^T(k)\chi \big(\omega(k)\big)\textbf{h}_a^*(k)\frac{dk}{2\pi}
\end{equation}

Using this formalism derived in \cite{Vanderveken}, the calculation of the spin-wave induced self and mutual impedances $Z_{\text{sw}}= i\omega L_\text{m}$ is possible using a simple analytic expression for the magnetic susceptibility if the dynamic Oersted field is known. Given the antenna’s minimal dimensions ($\leq\SI{1}{\micro\metre}$ thickness and width), a uniform current density can be assumed, while the spin-wave feedback on the current distribution is neglected. An analytical expression of the Oersted field of rectangular wires with uniform current distribution is derived from Biot-Savart's law in \cite{Chumakov}. By applying the superposition principle, the total antenna field is represented as the combined contributions from individual antenna conductors, each weighted by its corresponding current density. This simplified description effectively captures the field distribution for the system under investigation. The calculated impedances can then be converted to S-parameters using commonly known relations \cite{Pozar}.\\ 
Most microwave systems are impedance-matched; deviations from this value lead to reflections at the contacts, reducing transmission efficiency. For this study, the reference impedance is chosen as \SI{50}{\ohm}. Thus, effective coupling of energy into and out of the spin-wave system requires a real impedance component of \SI{50}{\ohm} to minimize reflections and maximize conversion efficiency. At resonance, the real part of the spin-wave impedance is maximized, as evident from Eq. (\ref{equ:SelfInductance}). The prefactor in Eqs. (\ref{equ:SelfInductance}) and (\ref{equ:MutualInductance}) further indicate that optimizing either the antenna length or the magnetic layer thickness enhances the impedance.\\
Including the electrical properties of the antennas introduces a series connection of the antenna impedance and the spin-wave impedance as outlined in \cite{RobertDesign}. From this analytic model, the following conclusions can be drawn:
\begin{enumerate}[itemsep=0.2cm,parsep=0.2cm]
    \item Adjusting the antenna length scales the total impedance but does not alter the proportion of power dissipated in the antenna versus the power coupled into the spin-wave system, as both scale linearly in antenna length.
    \item Increasing the magnetic layer thickness increases the spin-wave induced impedance, while keeping the antenna resistance unchanged. Therefore a larger fraction of power inserted into the antenna, is converted to spin-waves.
    \item For a given antenna, using a thicker magnetic layer is advantageous, as it reduces the relative impact of ohmic losses in the antenna as long as the dispersion does not level out within the excitation efficiency of the antenna.
\end{enumerate}
Additionally, the efficiency of spin-wave transducers is influenced by losses within the spin-wave system itself. In the Damon-Eshbach  configuration, both the decay length and group velocity of spin waves increase with film thickness, reducing propagation time and therefore damping-related losses.\\

\subsection{Numerical simulation}

Although the analytic model is insightful, it relies on approximations such as homogeneous field and mode profiles, plane-wave propagation, and a single-mode system. For thicker YIG films, the analytical description from \cite{Vanderveken} becomes inaccurate, and the assumption of uniform antenna current density further limits its precision. To overcome these limitations, we employ the simulation model from \cite{RobertDesign} to calculate both the spin-wave-induced and electrical impedances. In this approach, electromagnetic simulations first determine the antenna field distributions, which then serve as excitation fields in micromagnetic simulations to compute the resulting spin-wave excitations. By solving a lumped circuit representation, the spin-wave impedances are extracted from these micromagnetic simulations, inherently accounting for multiple modes with inhomogeneous profiles. Due to its greater accuracy, this simulation-based method is used in Section \ref{sec:Losses} to determine loss contributions, replacing the analytical approximation.

\section{Results \label{sec:Results}}
\subsection{Spin-wave transmission}\label{sec:Transmission}

\begin{figure}[h]
\includegraphics{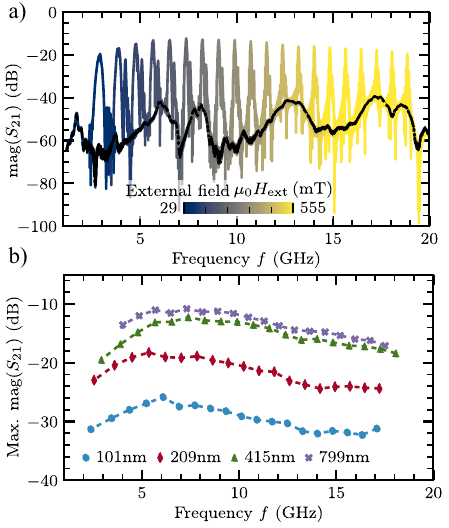}
\caption{Magnitude of the transmitted spin-wave signal. a) Transmission spectra for different external magnetic fields coded in color for the $\SI{415}{\nano\metre}$ thick YIG film. Indicated in black is the magnitude of the directly transmitted electric signal without spin-wave contribution. b) Extracted magnitudes of maximum spin-wave transmission signal for all YIG thicknesses tested.\label{fig:Transmission}}
\end{figure}

In the experiment, the entire S-matrix is acquired, enabling simultaneous analysis of the input antenna's reflection behavior ($S_{11}$) and spin-wave transmission ($S_{21}$). Figures \ref{fig:ExpSetup} c) and d) illustrate the measured and calculated reflection and transmission parameters for a representative field of $\SI{138.7}{\milli \tesla}$ on a $\SI{209}{\nano\metre}$ thick YIG film. According to Eq. (\ref{equ:SelfInductance}), the self impedance's frequency dependence is determined by the dynamic susceptibility frequency dependence weighted by the antenna field wave-vector distribution, that is shown in Fig \ref{fig:ExpSetup} b). The three maxima in antenna field wave-vector distribution consequently result in the three maxima of the reflection coefficient shown in Fig. \ref{fig:ExpSetup} c).\\
Using Eqs. (\ref{equ:SelfInductance}) and (\ref{equ:MutualInductance}), the expected reflection and transmission spectra are calculated, applying parameters derived from ferromagnetic resonance measurements on a reference film: $M_\text{s}=\SI{155.3}{\kilo\ampere\per\metre}$ and $\alpha=4.7\cdot 10^{-4}$. To account for the electrical characteristics of the antenna, off-resonant measurements of the experimental structure are included in series to the spin-wave impedances. The total impedances are then converted to scattering parameters using standard relations \cite{Pozar}.\\
The modeled and measured reflection and transmission spectra are shown in Figs. \ref{fig:ExpSetup} c) \-- e). Notably, the antenna excitation efficiency leads to three separated resonances in reflection $S_{11}$. Likewise, according transmission bands are observed in transmission $S_{21}$, with higher order peaks being less pronounced due to decreased excitation efficiency and group velocity.
In addition, outside of the spin-wave transmission bands a finite signal level is observed due to electromagnetic (EM) crosstalk between the antennas \cite{Bunea}.\\
Qualitative agreement between the modeled and measured S-parameters is generally good, though the measured data exhibit additional structural features. Consistent with observations in \cite{RobertDesign}, an extra peak at lower frequencies is attributed to the influence of contact pads and tapered lines, which can locally drive spin waves under a different angle to the external field at different resonance frequencies, that do not reach the output transducer. The oscillations towards higher frequencies are attributed to the interference between spin wavess and the EM crosstalk.\\
In the experiment, the operating frequency band is not fundamentally dictated by the CPW structure but by the spin-wave dispersion relation accessible to the wave-vector spectrum of the CPW. Adjusting the external field magnitude thus allows for tuning of the spin-wave transmission band in frequency. As shown in Fig. \ref{fig:Transmission}. a) for $\SI{415}{\nano\metre}$ thick YIG, the spin-wave transmission spectra remain consistent across the tested field range but become narrower at higher fields. This arises from the decreasing group velocity of the Damon-Eshbach mode, leading to  a compression of the dispersion relation into a narrower frequency band and increased propagation loss. Calculated and measured decay lengths and propagation losses are provided in the supplemental material.\\
Throughout the tested range, the maximum transmitted spin-wave signal remains more than $\SI{20}{dB}$ stronger than the electrically transmitted background. To track the evolution of spin-wave transmission, Fig. \ref{fig:Transmission} b) presents the maximum transmission magnitude at different external fields for all tested YIG thicknesses. The trend is consistent across samples, showing a clear correlation: thicker YIG films exhibit lower insertion loss, aligning with the conclusions of Section \ref{sec:Theory}. Notably, relative transmission magnitudes as high as $\SI{-11.5}{dB}$ are observed for the thickest YIG film, demonstrating the potential of micron-scale magnonic transmission lines also for sub-micrometer YIG thicknesses.\\
Please note that a further increase of the YIG thickness does not necessarily increase the spin-wave transmission for a given fixed antenna geometry due to changes of the dispersion relation impacting on the group velocity (see Fig. \ref{fig:ExpSetup} b). In addition, according to the analytic model of \cite{Vanderveken}, the spin-wave-induced impedance increases linearly with the YIG thickness (see Eq. (\ref{equ:SelfInductance})), which can lead to higher reflections. This illustrates the need for a comprehensive optimization strategy that requires a detailed characterization of the loss channels, which is provided in the following section.

\subsection{Loss channels \label{sec:Losses}}

There are multiple loss channels affecting power flow in the studied structures. In magnonic devices, e.g. one of the dominant loss mechanisms is high reflection loss due to impedance mismatch with the standard $\SI{50}{\ohm}$ microwave periphery. To systematically analyze these losses, the first step is to quantify the total power loss. So far, only transmitted power has been considered, given by

\begin{equation}\label{equ:TransFactor}
\epsilon_{\text{trans}} = \vert S_{21}\vert^2\cdot100\%.
\end{equation}

However, the  fraction of reflected power can be directly obtained from the experiment via the $S_{11}$ parameter as 

\begin{equation}\label{equ:ReflFactor}´´
\epsilon_{\text{refl}} = \vert S_{11}\vert^2\cdot100\%.
\end{equation}

The remaining power that is neither transmitted nor reflected is dissipated within the device is given by the dissipation factor $\epsilon_\text{diss}$ as

\begin{equation}\label{equ:DissFactor}
\epsilon_\text{diss} = (1-\vert S_{21}\vert^2 -\vert S_{11}\vert^2)\cdot100\%.
\end{equation}

\begin{figure*}[htb]
\includegraphics{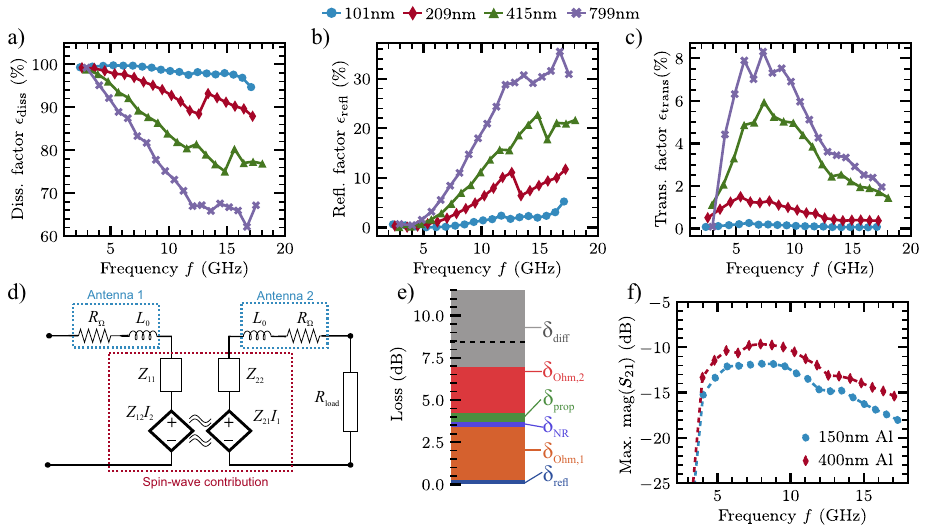}
\caption{Fraction of the supplied power, that is dissipated (a), reflected (b) or transmitted (c) at the point of maximum transmission for all tested YIG thicknesses and applied external fields. d) Equivalent circuit for the transducers adapted from \cite{RobertDesign} used for the quantification of loss channels in e). The difference to the measured insertion loss is marked in grey. A dashed line marks the simulated insertion loss. f) Extracted magnitudes of maximum spin-wave transmission signal for antennas of different thickness on $\SI{799}{nm}$ YIG. \label{fig:Losses}
}
\end{figure*}

Figure \ref{fig:Losses} a) \-- c) illustrate the relative fractions of dissipated, reflected and transmitted power for all tested YIG films and external magnetic fields, evaluated at the frequency of maximum spin-wave transmission.  In all cases, the power dissipation consistently exceeds $60\%$, reaching up to $90\%$ for thinner YIG films. Simultaneously, both the reflected and transmitted power components are significantly higher for thicker YIG layers. This indicates that despite higher reflection losses in thick YIG layers, the overall transmission is improved by reducing the power dissipation. This behavior can be understood in terms of the inductive nature of the spin-wave-induced impedance, as described by Eq. (\ref{equ:SelfInductance}), and the assumption that the electrical resistance of the antenna and the spin-wave impedance are in series (see Fig. \ref{fig:Losses} d)). As a result, a greater share of the injected power is converted through the spin-wave resistance rather than dissipated by the antenna Ohmic resistance. This desirable redistribution reduces the overall power dissipation, even as larger reflections lead to a decrease in the total power coupled into the device. However, even in these cases, the majority of the power is still lost.
The following section quantifies the dominant loss channels to determine their individual contributions.  For this purpose, the overall efficiency of the transducer is modeled as a product chain of the partial power transfer efficiencies of different components or parts of the overall system:

\begin{equation}\label{equ:Efficiencies}
\eta = \eta_{\text{refl}}\cdot \eta_{\text{Ohm,1}}\cdot\eta_{\text{nr}}\cdot\eta_{\text{prop}}\cdot\eta_{\text{Ohm,2}}
\end{equation}

This product can be written as the sum of all partial losses, following the conversion to dB,

\begin{equation}\label{equ:Losses}
\delta_{i} =-10\log_{10}\bigg(\frac{P_i}{P_{\text{available}}}\bigg) =  -10\log_{10}(\eta_{i})
\end{equation}

with the power transferred at this step $P_i$ and the power available at this step $P_{\text{available}}$.
The first term in the Eq.  (\ref{equ:Efficiencies}) accounts for the impedance mismatch and the resulting reflection losses. The terms $\eta_{\text{Ohm,1}}$ and $\eta_{\text{Ohm,2}}$ describe the power losses due to the Ohmic resistance of antennas 1 and 2, assuming a load resistance of $R_\text{load}=\SI{50}{\ohm}$ at the output of antenna 2. The corresponding equivalent circuit is shown in Fig. \ref{fig:Losses} d), adapted from \cite{RobertDesign}. Additionally, losses from spin waves propagating in the undesired direction $\eta_\text{nr}$ are considered, with a more detailed discussion provided in Section \ref{sec:Isolation}. Finally, the power losses due to spin-wave damping during propagation $\eta_\text{prop}$ are also estimated. This analysis is exemplarily conducted for the $\SI{799}{\nano\metre}$ thick YIG sample at an external field of $\SI{139}{\milli \tesla}$, specifically at the frequency of maximum transmission at around $\SI{6.4}{\giga\hertz}$. Based on these conditions, the individual loss contributions for this particular case are determined as follows:

\begin{equation}\label{equ:Reflection}
\eta_{\text{refl}} = 1 - \vert S_{11}\vert^2, \quad\delta_{\text{refl}}\approx \SI{0.25}{dB}
\end{equation}

\begin{equation}\label{equ:Ohm1}
\eta_{\text{Ohm,1}} = \frac{R_\text{sw}}{R_\text{sw}+R_\Omega}, \quad\delta_{\text{Ohm,1}}\approx \SI{3.14}{dB}
\end{equation}

\begin{equation}\label{equ:NR}
\eta_{\text{NR}} = \bigg(\frac{\vert S_{21}\vert}{\vert S_{21}\vert + \vert S_{12}\vert}\bigg)^2, \quad\delta_{\text{NR}}\approx \SI{0.3}{dB}
\end{equation}

\begin{equation}\label{equ:Prop}
\eta_{\text{prop}} = \text{exp}\Bigg(\frac{-2\cdot d}{l_{\text{sw}}}\Bigg), \quad\delta_{\text{prop}}\approx \SI{0.54}{dB}
\end{equation}

\begin{equation}\label{equ:Ohm2}
\eta_{\text{Ohm,2}} = \frac{R_\text{load}}{R_\text{load}+R_\Omega}, \quad\delta_{\text{Ohm,2}}\approx \SI{2.74}{dB}
\end{equation}

with the simulated spin-wave resistance $R_{\text{sw}}=\text{Re}(Z_{11})=\SI{43.95}{\ohm}$, the simulated Ohmic resistance of the antenna $R_{\Omega}=\SI{44.4}{\ohm}$, and the spin-wave amplitude decay length $l_{\text{sw}}=\SI{336}{\micro\metre}$, whose determination is detailed in the supplementary material. The spin-wave and antenna resistances required for calculating the Ohmic losses were numerically obtained using the simulation model from \cite{RobertDesign} for identical parameters.

The total quantified losses remain approximately $\delta_\text{diff} = \SI{4.6}{dB}$ lower than the insertion loss measured directly in the experiment (see Fig. \ref{fig:Losses} e)). This discrepancy may stem from several factors.
This difference may be attributed to several factors. The calculated total insertion loss in the simulation is around $\SI{8.4}{dB}$, which exceeds the sum of all individually quantified loss contributions. One potential cause is high spin wave intensity below the region of the excitation antenna in the simulation. In this region, nonlinear scattering processes may occur, acting as additional loss channels. In addition, the strong dynamic magnetization may locally distort the dispersion and prevent part of the excited spin waves from propagating out of the antenna region, contributing to further losses. Here, $\eta_{\text{Ohm,1}}$ only includes the part of the energy that was originally coupled into the magnetic system. However, some of this energy can be dissipated locally without contributing to the transmission of spin waves. Additional sources of loss that are not considered in the simulation include resistive losses in the contact lines and effects at the metal-YIG interface. In particular, the interface can lead to locally increased attenuation below the antenna structures, which further increases the losses due to spin wave dissipation or scattering.
Nevertheless, in direct comparison it becomes obvious, that in this reference case, the main source of losses inside of the transducer is caused by the high Ohmic resistance of the small current carrying lines of the antenna, amounting to $\SI{5.9}{dB}$ in total.

Thus, to further reduce insertion loss, it is beneficial to use antennas with lower resistance. To test this hypothesis, two additional devices were fabricated with antennas of identical design but made from aluminum instead of gold, primarily to reduce material costs. A comparison was conducted between antennas made from $\SI{150}{\nano\metre}$ and $\SI{400}{\nano\metre}$ thick aluminum. Applying the same loss quantification approach, changes in antenna resistance influence both reflection and Ohmic losses. Theoretical estimates predict an insertion loss reduction of approximately $\SI{2}{dB}$.

Measured resistance values, obtained from the off-resonant spectrum and an example spin-wave transmission spectrum, are provided in the supplementary material. Experimental results confirm an improvement of more than $\SI{2}{dB}$, depending on frequency, with insertion losses below $\SI{10}{dB}$ achieved using the thicker aluminum antennas, shown in Fig. \ref{fig:Losses} f). While further increasing antenna thickness could reduce Ohmic resistance even more, this would also lower spin-wave impedances, as a larger portion of the electric current would be located farther from the antenna. This dependence on antenna thickness is discussed in the supplementary material and has also been reported in the literature for out-of-plane magnetized YIG films \cite{Robbins}. 

Despite this trade-off, the ratio between spin-wave resistance and Ohmic resistance continues to increase within the tested parameter range, suggesting further potential for loss reduction. Ultimately, the findings demonstrate that the key to achieving efficient spin-wave transducers on the microscale, particularly for sub-micrometer-thick YIG films, is the use of low-resistance antennas.

\subsection{Non-reciprocity \label{sec:Isolation}}
\begin{figure*}[htb]
\includegraphics{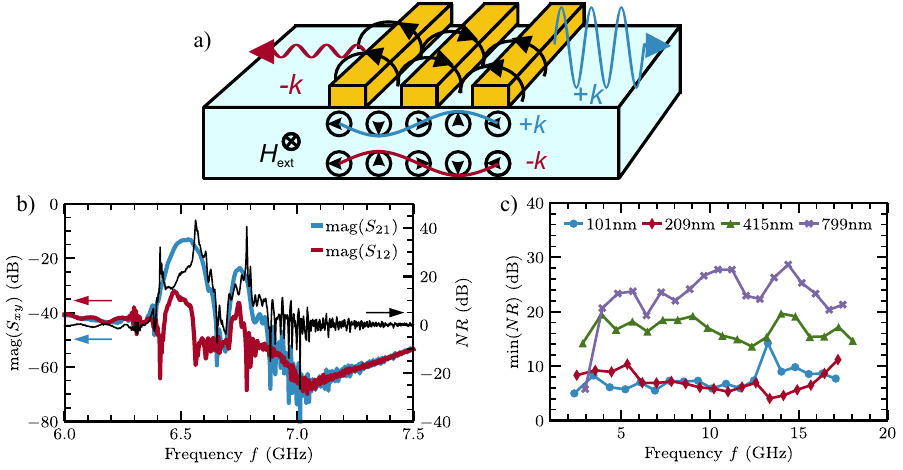}
\caption{a) Schematic illustration of spin-wave excitation in opposite propagation directions in Damon-Eshbach configuration. b) Measured insertion loss for opposite propagation directions in red and blue for the $\SI{415}{\nano\metre}$ YIG at an external field $\mu_0H_\text{ext}= \SI{142}{\milli \tesla}$. The ratio is shown in black. c) Minimum extracted ratios across the $\SI{3}{dB}$ bandwidths for all tested YIG thicknesses.\label{fig:NonReciprocity}
}
\end{figure*}

The experiments were conducted in the Damon-Eshbach configuration, which enhances spin-wave radiation resistance by leveraging both in-plane and out-of-plane components of the antenna’s excitation field \cite{Schneider}. This improves energy efficiency by utilizing the full field spectrum \cite{DevolderUni}. However, it also introduces a phase relationship between the generated spin waves, requiring alignment with the antenna field for constructive interference \cite{KostylevProp}. As shown schematically in Fig. \ref{fig:NonReciprocity} a), this results in spin-wave excitation in one direction while effectively suppressing it in the opposite direction.\\
In the experiment, this directional dependence of interference manifests as nonreciprocal signal transmission, a desired trait for many microwave applications. The measured transmission spectra confirm this, with one direction showing much lower insertion loss compared to the other, as anticipated from theoretical predictions and previous studies \cite{LiYi}. An example measurement for opposite propagation directions is shown in Fig. \ref{fig:NonReciprocity} b), along with the isolation $NR$ as the ratio of preferred and suppressed direction amplitudes.\\
The smaller bandwidth of the spin wave transmission observed in the suppressed direction arises from the lower amplitude, leading to a faster decay below the level of EM background signal. Towards the edges of both bands, a phase shift of $\pi$ between the spin wave and the electrical signals can cause significant destructive interference, resulting in sharp dips at the transmission band edges (Fig. \ref{fig:NonReciprocity} b)). To quantify isolation, we extract the minimum isolation value within the $\SI{3}{dB}$-bandwidth around the peak transmission for all external magnetic fields and YIG film thicknesses, as this defines the lower bound of achievable isolation in a potential device.\\
The extracted values, shown in Fig. \ref{fig:NonReciprocity} c), reveal that isolation does not follow a uniform trend across all measured thicknesses as a function of the external field. This variation likely stems from finite difference in slope between both propagation directions near the maximum of $S_{21}$ in Fig. \ref{fig:NonReciprocity} b). However, a clear correlation with film thickness is observed: isolation increases sharply with thickness, reaching up to $\SI{30}{dB}$ over the $\SI{3}{dB}$ bandwidth. This enhanced non-reciprocity is attributed to the improved matching between the spin precession ellipticity and the antenna field ellipticity, maximizing interference modulation. As demonstrated in \cite{LiYi}, optimal isolation occurs when the width of the electrical lines equals the YIG layer thickness. Given the $\SI{1}{\micro\metre}$ wide lines used in this study, the highest isolation is expected for the thickest YIG film of $\SI{799}{\nano\metre}$. These findings highlight the potential of microstructured transmission lines for compact, high-performance RF isolators.

\subsection{Power dependence \label{sec:Power}}

\begin{figure}[h]
\includegraphics{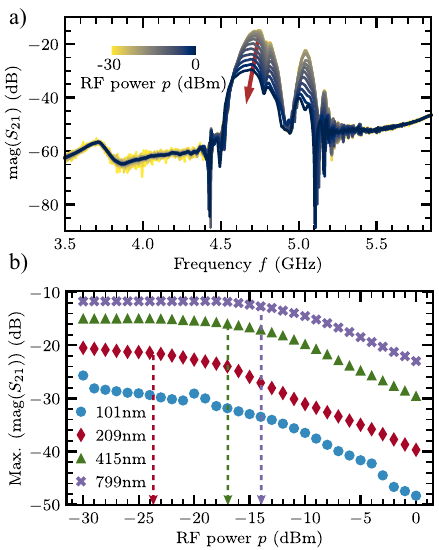}
\caption{a) Measured transmission magnitude for a $\SI{415}{\nano\metre}$ thick YIG film at an external field $\mu_0H_\text{ext}=\SI{84}{\milli \tesla}$ for varying input power. Due to the measurement of transmission magnitude relative to the input power, at higher power the transmission appears to decrease. b) Extracted maximum transmission for different input powers and YIG film thicknesses. The $\SI{1}{dB}$ compression points are marked for the thicker samples, while the $\SI{101}{\nano\metre}$ sample is nonlinear across the whole tested range. \label{fig:PowerDependence}}
\end{figure}

In the previous sections, all experiments were conducted with a constant input RF power of $P=\SI{-25}{dBm}$. In the linear regime, no change in transmission amplitude is expected, as the VNA measures only the relative transmitted power compared to the input power. However, when the input power exceeds a threshold, nonlinear excitations introduce additional loss channels in the magnonic system, reducing the relative transmission. The results of power dependent transmission for the $\SI{415}{\nano\metre}$ thick YIG sample, shown in Fig. \ref{fig:PowerDependence} a), indicate that at lower power levels, the spectra indeed remain unchanged. However, when the power exceeds a certain threshold, a noticeable reduction in transmission is observed across the entire spectrum. This reduction is due to nonlinear spin-wave processes introducing additional losses. Furthermore, a shift of the transmission maximum to lower frequencies is observed, as the effective magnetization decreases when the precession amplitude increases, lowering the resonance frequency at a fixed wave-vector \cite{Krivosik}. 

The power-dependent performance is evaluated for all YIG film thicknesses, as shown in Fig. \ref{fig:PowerDependence} b), which presents the maximum relative transmission for each applied input power. A clear trend emerges: the onset of transmission degradation shifts to higher power levels with increasing YIG thickness, as can be seen from the indicated $\SI{1}{dB}$ compression points. Notably, the $\SI{101}{\nano\metre}$ film shows a drop already at the lowest tested power. The shift in $\SI{1}{dB}$ compression point is attributed to the higher group velocity in thicker films as well as the larger magnetic volume itself, which reduces energy density in the YIG at the same input power. The magnitude of this shift aligns well with the expected $\SI{3}{dB}$ per thickness doubling \cite{Davidkova}, but for thinner films, the shift is larger than expected, likely due to the lower group velocity and higher localized energy accumulation near the antenna.

\subsection{Permanent magnet integration}

\begin{figure}[h]
\includegraphics{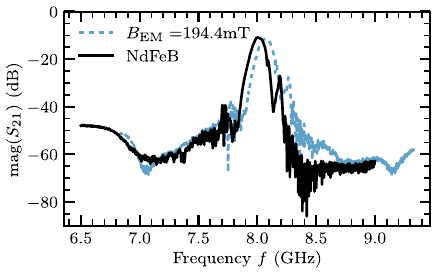}
\caption{Measured transmission spectrum using a permanent magnet alongside a reference measurement using an electromagnet for the $\SI{799}{\nano\metre}$ YIG film. \label{fig:PermanentMagnet}}
\end{figure}

To assess the feasibility of using a small permanent magnet instead of an electromagnet for future on-chip device integration, a $3\times 3 \times 3 \, \si{\milli\metre\cubed}$ NdFeB magnetic cube is centered below the device under test (DUT), providing a bias field in DE configuration. Following a SOLT calibration, the spin-wave transmission spectrum is acquired, presented in Fig. \ref{fig:PermanentMagnet}.\\
A comparison of the spin-wave transmission spectra obtained with the electromagnet and the permanent magnet reveals that both the spectral shape and insertion loss characteristics are nearly identical. This demonstrates that using a small permanent magnet for biasing does not compromise device performance. The microscale dimensions of the DUT facilitate seamless integration of small permanent magnets, as their size remains significantly larger than the device active area, minimizing the impact of field gradients and inhomogeneities. Moreover, the observed spectra correspond to a magnetic field strength of approximately $\SI{190}{\milli \tesla}$, enabling fully passive operation within the 8 GHz frequency band.\\
This result not only confirms the feasibility of integrating small magnets directly onto chips but also enhances the system’s overall energy efficiency. By eliminating the need for an external power source to drive an electromagnet, the device operates in a fully passive manner, significantly reducing power consumption and improving the practicality of standalone, fully integrated systems for future applications.\\

\section{Conclusion}
In this work, we presented a study on microscale spin-wave transducers based on submicrometer YIG films. 
By identifying ohmic losses as the dominant loss channel rather than intrinsic spin-wave attenuation or reflection losses, we were able to achieve transmission levels of more than $\SI{-10}{dB}$. This underlines the potential for further improvements through electrical optimization. In addition, the devices intrinsically provide passive isolation properties with isolations up to $\SI{30}{dB}$, a favorable trait for many rf-applications. 
Furthermore, the observed power dependence underlines the ease with which nonlinear behavior can be achieved
in magnetic devices. This opens up possibilities for advanced features such as nonlinear limiters or neural networks that utilize efficient microwave-to-spin-wave conversion.
Finally, integration with small permanent magnets demonstrates a practical path toward compact, bias-field-free magnetic components, enabling scalable implementation in RF and communication technologies.

\section{Acknowledgements}
Financial support by the EU Horizon Europe research and innovation program within the projects “MandMEMS” (Grant No. 101070536)  and “SPIDER” (Grant No. 101070417) is gratefully acknowledged. 
\bibliography{refs}
\end{document}